# Government spending and multi-category treatment effects: The modified conditional independence assumption

Koiti Yano[a][*]

First Draft: July 17, 2020

[a] Komazawa University, 1-23-1Komazawa, Setagaya-ku, Tokyo, 154-8525, Japan (e-mail: koiti@komazawa-u.ac.jp).
[*]corresponding author

Abstract: I devise a novel approach to evaluate the effectiveness of fiscal policy in the short run with multi-category treatment effects and inverse probability weighting based on the potential outcome framework. This study's main contribution to the literature is the proposed modified conditional independence assumption to improve the evaluation of fiscal policy. Using this approach, I analyze the effects of government spending on the US economy from 1992 to 2019. The empirical study indicates that large fiscal contraction generates a negative effect on the economic growth rate, and small and large fiscal expansions realize a positive effect. However, these effects are not significant in the traditional multiple regression approach. I conclude that this new approach significantly improves the evaluation of fiscal policy.

## 1. Introduction

Angrist et al. (2017), Angrist and Kuersteiner (2011), and Jordà and Taylor (2016) develop a novel approach for the evaluation of macroeconomic policy interventions. Based on Rosenbaum and Rubin (1983), their approach is a flexible semiparametric time series method for the estimation of the causal effect of economic policy on macroeconomic variables. The advantage of this approach is that assumptions about the process generating macroeconomic variables are not necessary.

However, there is room for improvement in their approach in terms of evaluating the effectiveness of fiscal policy in the short run because their standard form of the conditional independent assumption (CIA) is violated in the fiscal policy. Therefore, I devise a new approach, applying multi-category treatment effects and inverse probability weighting (IPW), with modifications of previous studies.[1] The main contribution of this study is that I modify the standard CIA to fit the fiscal policy. I stress that my modification is naturally required because government spending is included in the output of the economy. For example, government spending is a determining factor of a country's gross domestic product (GDP).

In this study, I focus on the effects of government spending. Angrist and Kuersteiner (2011) and Angrist et al. (2017) focus on monetary policy, and Jordà and Taylor (2016) focus on fiscal consolidation. Peren Arin et al. (2015), however, argue that there has been a revival of interest in the relation between government spending and economic growth, and not only in fiscal consolidation after the global financial collapse in 2008. Therefore, I focus on government spending of the US fiscal policy from 1992 to 2019.

Batini et al. (2014) review studies on fiscal policy analysis and argue that there are two common approaches: dynamic stochastic general equilibrium models (DSGEs) and structural vector autoregression models (SVARs). Both approaches depend on specified macroeconomic models. Angrist et al. (2017) point out that their validity depends on how accurately the macroeconomic models describe the economy. An alternative strategy to DSGEs and SVARs is the narrative approach proposed by Romer and Romer (1989). It does not depend on specified macroeconomic models, but the subjective manner in which shocks are identified is a drawback. My novel approach, based on Angrist et al. (2017) and related papers, avoids this drawback.

The remainder of this paper is organized as follows. In section 2, I describe my novel method and the modified conditional independence assumption (MCIA). In section 3, the results of the empirical study on US fiscal policy are presented, and the conclusions and discussion are described in section 4.

---

[1] My approach is based on a simplified version of Angrist and Kuersteiner (2011), Angrist et al. (2017), and Jordà and Taylor (2016) to focus my modification of the CIA.

## 2. Method

Similar to Angrist et al. (2017), I adopt the potential outcome framework in the context of multi-category treatment effects. The economy is assigned one of $J$ possible levels $(1, \cdots, j, \cdots, J)$ of fiscal policy in time period $t = 1, \cdots, T$. For each time period $t$, the random vector $z_t = (y_t, q_t, x_t', z_t')'$ is observed, where $y_t$ is the growth rate of output, $q_t \in (1, \cdots, J)$ is the level of the growth rate of government spending, $x_t$ is a $k \times 1$ vector of covariates, and $z_t$ is a $l \times 1$ vector of macroeconomic variables. In the framework, the observed growth rate $y_t$ is assumed to satisfy the equation below.

$$y_t = d_t(1)y_t(1) + \cdots + d_t(J)y_t(J),$$

where $d_t(j) = \mathbf{1}(q_t = j)$, $y_t(1), \cdots, y_t(J)$ are $J$ potential outcomes, and $\mathbf{1}(\cdot)$ is the indicator function.

Angrist et al. (2017) adopt the standard CIA and the no-empty-cell assumption as follows.
  Assumption 1. For all $t = 1, \cdots, T$:
  (a) (CIA) $[y_{t+h}(j) - y_t] \perp d_t(j) | x_{t-k}$, for all $h > 0$, and $k \geq 0$,
  (b) (no-empty-cell) $0 < e_{min} < p_j(x_t)$, with $p_j(x_t) = \mathbb{P}(q_t = j | x_t)$,
where $p_j(x_t)$ is the generalized propensity score proposed by Imbens (2000).

Although assumption 1(a) is standard in the causal model, which is proposed by Rosenbaum and Rubin (1983), it is not suitable for fiscal policy evaluation. Even if $x_{t-k}$ is given, $[y_{t+h}(j) - y_t]$ is not independent on $d_t(j)$ because $y_t$ includes the growth rate of government spending $g_t$, and $d_t(j)$ is determined by $g_t$. Thus, in the fiscal policy evaluation, assumption 1(a) is violated. Assumption 1(b) states that for every possible $x_t$ in the population, there is a strictly positive probability that the economy with that covariate pattern could be assigned to each treatment level. This ensures that the generalized propensity score is bounded away from zero. This is an important condition for a semiparametric efficient estimation.

The main contribution of this study is that I modify the standard CIA to avoid its violation for fiscal policy evaluation. The MCIA and the standard no-empty-cell assumption is defined as follows:
  Assumption 2. For all $j = 1, \cdots, J$:
  (a) (MCIA) $[y_{t+h}(j) - \hat{y}_t] \perp d_t(j) | x_{t-l}$, for all $h > 0$, and $l \geq 1$,
  (b) (no-empty-cell) $0 < e_{min} < p_j(x_t)$, with $p_j(x_t) = \mathbb{P}(w = j | x_t)$,
where $\hat{y}_t$ is a forecasted value of $y_t$ based on a vector of macroeconomic variables, $z_{t-1}$. $\hat{y}_t$ is independent on $d_t(j)$ and $g_t$ because it is a forecasted value based on $z_{t-1}$. Consequently, assumption 2(a) is not violated in the fiscal policy evaluation. Assumption 2(b) is unchanged from 1(b).

The estimand of interest is the improvement of growth rates, which are given by $\boldsymbol{\mu} = (\mu_1, \cdots, \mu_J)'$ with $\mu_j = E[y_{t+h}(j) - \hat{y}_t]$. When $q_t = j$, assumption 2(a) implies that

$$E\left[\frac{d_{t+h}(j)y_{t+h}(j) - \hat{y}_t}{p_j(\boldsymbol{x}_{t-l})}\right] = E\left[E\left[\frac{d_{t+h}(j)y_{t+h}(j)}{p_j(\boldsymbol{x}_{t-l})}\bigg|\boldsymbol{x}_{t-l}\right]\right] - \hat{y}_t$$

$$= E\left[E\left[\frac{1}{p_j(\boldsymbol{x}_{t-l})}\bigg|\boldsymbol{x}_{t-l}\right]E[d_{t+h}(j)|\boldsymbol{x}_{t-l}]E[y_{t+h}(j)|\boldsymbol{x}_{t-l}]\right] - \hat{y}_t = E[E[y_{t+h}(j)|\boldsymbol{x}_{t-l}]] - \hat{y}_t$$

$$= E[y_{t+h}(j) - \hat{y}_t] = \mu_j.$$

Following Angrist et al. (2017), I adopt IPW to estimate the estimand. For brevity, I restrict $l$ to unity in this study.

My approach is summarized below. The growth rate of government spending is classified into four groups: large fiscal contraction, small fiscal contraction, small fiscal expansion, and large fiscal expansion. [2] For these groups, I use dummy variables that take the value of 0 or 1 to indicate the absence or presence of a categorical effect, respectively. The dependent variable $[y_{t+h}(j) - \hat{y}_t]$ is calculated, where $y_t$ is the log of real GDP (RGDP), and $\hat{y}_t$ is an estimate of $y_t$ based on the vector of quarterly macroeconomic variables, $\boldsymbol{z}_{t-1}$. I estimate weights, which is the inverse of the generalized propensity score, based on $\boldsymbol{x}_t$.[3] Following Angrist et al. (2017), I finally estimate equation (1) with weighted least squares (WLS) based on the weights.

$$[y_{t+h}(j) - \hat{y}_t] = \beta_1 f_1 + \beta_2 f_2 + \beta_3 f_3 + \beta_4 f_4, \qquad (1)$$

where $f_1, f_2, f_3$, and $f_4$ are the dummy variables for the four groups of government spending and $\beta_1, \beta_2, \beta_3$, and $\beta_4$ are the estimators of the estimands, respectively. For brevity, I restrict $h$ to unity.

### 3. Empirical study

In this study, I focus on the relation between government spending and the growth of the US economy because there has been a revival of interest in this relation after the global financial collapse in 2008. In this section, estimating equation (1), I conduct an empirical study on the effects of government spending on the US economy using quarterly time series as covariates and macroeconomic variables from 1992 to 2019.[4] The log of output $y_t$ is the log of RGDP, and $\hat{y}_t$ is an estimate of $y_t$ based on $\boldsymbol{z}_{t-1}$.[5] The covariates, $\boldsymbol{x}_{t-1}$, are the log-difference of RGDP, the log-difference of commodity index, and the difference in unemployment rate at time period $t-1$.

---

[2] Suppose that $\sigma$ is the standard deviation of $g_t$. Large fiscal contraction is $g_t \leq -\sigma$, small fiscal contraction is $-\sigma < g_t \leq 0$, small fiscal expansion is $0 < g_t \leq \sigma$, and large fiscal expansion is $\sigma < g_t$.

[3] Details of the generalized propensity score are described in Imbens (2000).

[4] Codes for my empirical study are provided at the following URL: https://github.com/koiti-yano/gov_spend_and_multi

[5] The macroeconomic variables are RGDP, TED spread, commodity index, and unemployment rate. RGDP, commodity index; unemployment rates are log-transformed; and RGDP and unemployment rate are seasonally adjusted.

Table 1 reports the results for equation (1). The first column, labeled WLS(A2), contains the results of the baseline model based on my novel approach. To compare my approach and that with non-weighted ordinary least squares, I show the results of the OLS model in the second column, labeled OLS(A2). The third column, labeled WLS(A1), contains the results of the WLS model based on assumption 1 to compare assumptions 1 and 2.

The results of the baseline model are shown in the first column. They indicate that the coefficients of large fiscal contraction, small fiscal expansion, and large fiscal expansion are significant at the five percent level. Additionally, they show that large fiscal contraction generates a negative effect on the growth rate of the economy, and small fiscal expansion and large fiscal expansion realize a positive effect on the growth rate because the coefficients are significant at the five percent level, whereas the results of the second column indicate that the effects of government spending on the growth rate are not significant at the five percent level. The comparison shows that adopting IPW and assumption 2 improves the estimation. The third column reports the WLS estimates based on assumption 1. The estimates in the third column are greater than those in the first column. They indicate that adopting assumption 1 overestimates the effects of government spending. The results presented in Table 1 show that my approach improves the evaluation of fiscal policy.[6]

---

[6] I estimate equation (1) with maximum likelihood, and my conclusions are unchanged.

TABLE 1
REGRESSION RESULTS FOR $[y_{t+h}(j) - \hat{y}_t]$

| Independent variable | WLS (A2) | OLS (A2) | WLS (A1) |
|---|---|---|---|
| Large fiscal contraction (Intercept) | -0.0028 ** | -0.0014 | 0.0025 * |
|  | (0.0010) | (0.0010) | (0.0012) |
| Small fiscal contraction | 0.0022 | 0.0008 | 0.0028 |
|  | (0.0014) | (0.0013) | (0.0017) |
| Small fiscal expansion | 0.0032 * | 0.0020 | 0.0051 ** |
|  | (0.0014) | (0.0016) | (0.0018) |
| Large fiscal expansion | 0.0035 * | 0.0023 | 0.0049 ** |
|  | (0.0014) | (0.0012) | (0.0017) |
| Number of observations | 109 | 109 | 109 |
| $R^2$ | 0.0655 | 0.0386 | 0.1001 |

*Notes:* Table 1 reports the results for equation (1). The results of the baseline model are shown in the first column. They show that the coefficients of large fiscal contraction, small fiscal expansion, and large fiscal expansion are significant at the 5 percent level. They indicate the effectiveness of fiscal policy. Standard errors are in parentheses. *** Significant at the 0.1 percent level. ** Significant at the 1 percent level. * Significant at the 5 percent level.

*Source:* Author's calculations.

## 4. Conclusions

In this study, I devise a novel approach based on Angrist and Kuersteiner (2011), Angrist et al. (2017), and Jordà and Taylor (2016) by modifying the CIA to improve the evaluation of fiscal policy. Using this approach, I analyze the fiscal policy of the US economy. The empirical study indicates that large fiscal contraction generates a negative effect on the growth rate of the economy, and small and large fiscal expansions realize a positive effect on the growth rate because the coefficients of my regression are significant at the five percent level. By contrast, the results of the traditional multiple regression approach indicate that the effects of government spending on the growth rate are not significant at the same level. Additionally, the coefficients of the approach based on the standard CIA are greater than the estimates based on my approach. They indicate that adopting the standard CIA overestimates the effects of fiscal policy. I conclude that my novel approach significantly improves the evaluation of fiscal policy.

To focus on my modified CIA and keep this paper concise, I adopt a simplified version of the approach proposed in previous studies and restrict $h$ =1 in the empirical study. In another study, I extend this method to a more general form and $h$ > 1. Additionally, I analyze the fiscal policy of the Organisation for Economic Co-operation and Development countries for international comparison and provide clear political implications.

**Acknowledgments**

The author is grateful to Tae Okada for her helpful comments.

**Funding**

This research did not receive any specific grant from funding agencies in the public, commercial, or not-for-profit sectors.

**Declaration of interest**

Declarations of interest: none